\documentstyle[12pt,epsfig]{article}
\def\be{\begin{equation}} \def\ee{\end{equation}} \def\bea{\begin{eqnarray}}
\def\eea{\end{eqnarray}} \def\nnb{\nonumber}

\newcommand{\Slash}[1]{\ooalign{\hfil/\hfil\crcr$#1$}}
\newcommand{\eqn}[1]{\label{eq:#1}}
\newcommand{\refeq}[1]{(\ref{eq:#1})}

\newcommand{\Eq}{Eq.~\refeq}
\newcommand{\Eqs}[2]{Eqs.~(\ref{eq:#1}) and (\ref{eq:#2})}

\newcommand{\vecbox}[1]{\mbox{\boldmath{$#1$}}}

\def\piless{EFT($\Slash{\pi}$)}

\def\pislashx{ {\pi\hskip-0.45em /} }

\def\nopix{ EFT(\pislashx) }
\def\veft{$V_{EFT}$}
\def\vph{$V_{ph}$}

\def\vpiII{$V_{EFT(\pi)}^{II}$}
\def\vpilessII{$V_{\nopix}^{II}$}
\def\vpi{$V_{EFT(\pi)}$}
\def\vpiless{$V_{\nopix}$}

\def\pieft{EFT($\pi$)}

\def\eftpi{EFT($\pi$)}

\def\1s0{{}^1S_0}
\def\ts1{{}^3S_1}
\def\td1{{}^3D_1}
\def\ltap{\ \raise.3ex\hbox{$<$\kern-.75em\lower1ex\hbox{$\sim$}}\ }
\def\gtap{\ \raise.3ex\hbox{$>$\kern-.75em\lower1ex\hbox{$\sim$}}\ }
\def\ket#1{\vert#1\rangle}
\def\bra#1{\langle#1\vert}

\def\braketa#1{\langle#1\rangle}

\begin{document}

\hfill{TRI-PP-06-08}

\vskip 7mm

\begin{center}
{\Large \bf
Renormalization group analysis\\
of nuclear current operators 
}

\vskip 7mm

{\large Satoshi X. Nakamura$^{a,}$\footnote{Email:snakamura@triumf.ca}
and Shung-ichi Ando$^{b,}$\footnote{Email:sando@meson.skku.ac.kr}}

\vskip 7mm

{\large \it $^a$Theory Group, TRIUMF, 
4004 Wesbrook Mall, Vancouver, BC V6T 2A3, Canada}

{\large \it $^b$Department of Physics, Sungkyunkwan University,
Suwon, 440-746 Korea}

\end{center} 

\vskip 7mm

A Wilsonian renormalization group (WRG) equation
for nuclear current operators
in two-nucleon systems
is derived.
Nuclear current operators relevant to low-energy Gamow-Teller transitions
are analyzed using the WRG equation.
We employ the axial two-body current operators
from phenomenological models 
and heavy-baryon chiral perturbation theory,
which are quite different from one another 
in describing small scale physics.
After reducing the model space of the operators
using the WRG equation, we find that
there still remains a significant model dependence
at $\Lambda = 200$ MeV, where $\Lambda$ is the sharp cutoff
specifying the size of the model space.
A model independent effective current operator is found
at a rather small cutoff value, $\Lambda = 70$ MeV.
By simulating the effective current operator at $\Lambda=70$ MeV,
we obtain a current operator based on a pionless theory, 
thereby arguing that an equivalence relation exists 
between nuclear current operators 
of phenomenological models and those of effective field theories.

\vskip 5mm \noindent
PACS(s): 05.10.Cc, 25.10.+s.

\newpage
\noindent
{\bf 1. Introduction} 

A theoretical framework, based on the idea of effective field theory,
for describing nuclear system was proposed at the beginning of 1990's 
by Weinberg~\cite{weinberg}.
The framework provides us with
a systematic scheme for constructing nuclear operators 
such as nuclear forces and nuclear electroweak currents 
from an effective Lagrangian of the underlying theory. 
We will refer to it as nuclear effective field theory (NEFT).
For reviews, see, {\it e.g.}, 
Refs.~\cite{l-97,betal-00,bvk-arnps02,kp-arnps04,e-05}
and references therein.
Meanwhile, phenomenological models for the nuclear operators
have been conventionally used in nuclear physics.
Expressions and behaviors of those nuclear operators 
are quite different among the phenomenological models and NEFT,
whereas 
all of the nuclear operators give essentially the same reaction rates
for low-energy reactions in few-nucleon system,
{\it e.g.}, solar-neutrino reactions on the 
deuteron\cite{nsgk,netal,bck,ando}.
This implies that there exists an equivalence relation 
between the nuclear operators of phenomenological models and NEFT,
and thus it is interesting to seek for the relation. 

Recently, one of the authors (SXN) argued the relation
for the nuclear forces from a viewpoint of the renormalization group
(RG)~\cite{rg1}.
In that work, the author formulated 
a ``scenario'', which we describe in detail in the followings, 
for the relation between
the nuclear forces based on NEFT (\veft) and phenomenological nuclear
forces (\vph) paying attention to a difference in the size of the model
space between \veft\ and \vph.
One may construct many different \vph\ all of which reproduce 
the low-energy $NN$-data. 
Those nuclear forces are different 
from one another in modeling small scale phenomena.
As high-momentum states of the nucleon are integrated out and thus 
the model space of \vph\ is reduced,
information relevant to the {\it details} of
the small scale physics are gradually lost.
As a result, all \vph\ eventually converge to a single nuclear force
defined in a sufficiently small model space;
the model dependence found in the original \vph\ disappears.\footnote{
A unique low-momentum effective $NN$ potential 
with the cutoff value $\Lambda=2.1$ fm$^{-1}$,
known as $V_{low\mbox{-}k}$,
is derived using the Lee-Suzuki method 
by Bogner {\it et al.}\cite{vlowk}. 
}
The short-range part of the single low-momentum interaction
is accurately parameterized with the use of simple contact interactions,
which is the way NEFT describes the small scale physics.
After all, the parameterization of the single low-momentum interaction
is nothing more than \veft.
By construction, \veft\ obtained in this way does not have a dependence on
modeling the small scale physics.
This is the scenario for the relation among \veft\ and \vph.
In Ref.~\cite{rg1}, the author employed a Wilsonian renormalization
group (WRG) equation
derived by Birse, McGovern and Richardson~\cite{birse}\footnote{
The WRG equation derived by Birse {\it et al.} is
equivalent to the
Bloch-Horowitz method\cite{bh-np58}.
}
for the model space reduction,
and showed a result to convincingly argue
that the scenario for the
relation between \vph\ and \veft\ is realized.

It
would be interesting to extend 
the WRG analysis to the study of 
nuclear electroweak currents.
This is the main subject of this paper.
Similar to the nuclear forces, the expression and behavior 
of a nuclear exchange current operator of a phenomenological 
model are quite different from those of another model or NEFT,
and thus the reaction mechanism of 
an electroweak process is also model dependent.
The model dependence stems from modeling small scale physics. 
Thus,
if one reduces the model spaces of the different nuclear current operators,
it is expected that all the operators evolve to be essentially a single
nuclear current, which may be accurately simulated by the NEFT-based
parameterization. In this way, nuclear electroweak currents from
phenomenological models and from NEFT are related through RG.

In this work, we
derive a WRG equation
of nuclear current operator for two-nucleon system\footnote{
We will restrict ourselves to two-nucleon system in this work,
and thus consider neither the genuine three (or more)-body currents 
nor the generation of them 
due to the model-space reduction.}.
By using the WRG equation, 
we reduce the model space of nuclear current operators 
of either phenomenological models or
NEFT with the pion, and examine the evolution of the operators.
We are specifically concerned with the axial vector current 
associated with
the Gamow-Teller transition in the low-energy neutral-current neutrino
reaction on the deuteron ($\nu d \rightarrow \nu pn$),
where
our analysis can be simplified due to the absence of the Coulomb
interaction.
Finally, the effective operator, obtained as a result of the model-space
reduction, is simulated by a NEFT-based operator 
without the pion~\cite{crs-npa99}.

Here, we mention a choice of the model-space reduction scheme.
So far, some methods have been used to derive a low-momentum $NN$
interaction from phenomenological $NN$ interactions.
Those methods are the WRG method\cite{rg1,birse},
the Lee-Suzuki method\cite{vlowk,ls-plb80}, 
and the unitary transformation method\cite{ut}.
For detailed comparison of them, see Ref.~\cite{j-el05}.
In Ref.~\cite{rg1}, the author
argued that the use of the WRG method is 
consistent with the construction of the effective Lagrangian in NEFT.
This argument is based on that
an effective Lagrangian in NEFT is in principle 
obtained by integrating out high-energy degrees of freedom within 
the path integral formalism, and that solving the WRG equation is equivalent 
to performing a path integral.
Also in the reference, the three model-space reduction schemes are
applied to the nuclear forces and 
it is shown that only the WRG method generates an effective low-momentum
interaction which is accurately simulated by the NEFT-based
parameterization of the nuclear forces, 
even in a case of a small model
space relevant to a pionless EFT [\piless].\footnote{
An effective interaction obtained with the WRG method is on-shell energy
dependent. 
The importance of considering the on-shell energy dependence of a
NEFT-based nuclear force has been examined in Ref.~\cite{rg1,harada}.
}
Therefore, 
we regard the WRG method as being the most appropriate for studying the
relation between the models and NEFT, and
we adopt it as the model-space reduction method in this work.

This paper is organized as follows.
In Sec.~2, 
we derive the WRG equation of the 
nuclear current operators for two-nucleon system.
In Sec.~3,
we present the current operators employed in this work, 
and give a detailed description of our numerical analysis.
In Sec.~4, we show results of the numerical calculation
and give discussion on our results.
Finally, we give a concluding remark in Sec.~5.
In appendices, we give a detailed derivation of the WRG equation, a
discussion of a conservation law for an effective current operator, and
a discussion on a role of the tensor nuclear force in a pionless theory.

\vskip 2mm \noindent
{\bf 2 Wilsonian renormalization group equation 
for nuclear current operator}

We briefly discuss a Wilsonian renormalization group (WRG) equation 
for nuclear current operators.
A detailed derivation of the WRG equation is given  
in Appendix A. 
We start with a matrix element of a current operator 
in a two-nucleon system evaluated 
in the momentum space,
\begin{eqnarray}
\eqn{o_me}
\braketa{O} 
= \int_0^\Lambda\!\!\! k^2 dk \int_0^\Lambda\!\!\! k'^2 dk'\;
\psi_{p'}^{(\beta)*}(k')\; O^{(\beta,\alpha)}(k',k)\; \psi_p^{(\alpha)}(k) \ ,
\end{eqnarray}
where $\psi_p^{(\alpha)}(k)$ ($\psi_{p'}^{(\beta)}(k')$)
is the radial part of the wave function for 
the initial (final) two-nucleon state,
which are derived from an effective interaction with 
a cutoff $\Lambda$
(see \Eq{eq_wave} in Appendix A).
The quantity $p$ ($p'$) is an on-shell relative momentum for 
the initial (final) two-nucleon state, $p\equiv\sqrt{ME}$
($p'\equiv\sqrt{ME'}$) with $E$ ($E'$) being
the energy for the relative motion of the two nucleons
and $M$ is the nucleon mass,
and $\alpha$ ($\beta$) specifies a partial wave.
The radial part of the current operator between 
the $\beta$ and $\alpha$ partial waves are denoted by
$O^{(\beta,\alpha)}(k',k)$.  
The quantities 
$k$ and $k'$ are the magnitudes of 
the relative off-shell momenta of the two-nucleon system.
The cutoff $\Lambda$, which is the maximum magnitude 
of the relative momenta, specifies the size of 
the model space spanned by plane wave states
of the two-nucleon system.

\begin{figure}[t]
\begin{center}
\includegraphics[width=80mm]{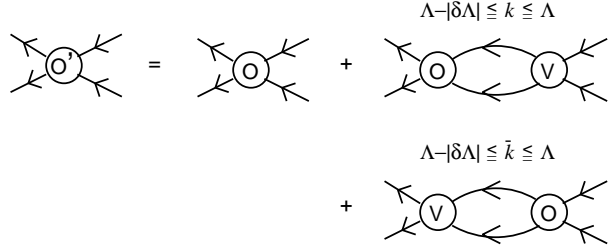}
\caption{\label{fig_rge} 
A graphical representation 
of the WRG equation for a current operator.
$O$ ($O'$) is an original (renormalized) current operator
and $V$ is a $NN$-interaction. 
In the loops, the magnitude of the relative momentum of the two nucleons 
is denoted by $\bar{k}$, which lies in the momentum shell 
that is integrated out. }
\end{center}
\end{figure}

We differentiate the both sides of \Eq{o_me}
and impose a condition that 
the matrix element is invariant with respect to changes in
$\Lambda$, {\it i.e.}, $d\braketa{O}/d\Lambda=0$.
This leads to
the WRG equation for the {\it effective} current operator,\footnote{
\label{footnote;path_int}
In Ref.~\cite{rg1},
the WRG equation for the nuclear force is derived using the path integral
method.
The WRG equation for the current operator, \Eq{rge}, is also derived
in essentially the same way.
}
\begin{eqnarray}
\eqn{rge}
&&
{\partial O^{(\beta,\alpha)}_\Lambda(k',k;p',p) \over\partial\Lambda}
\nonumber\\ 
&& = 
{M\over 2\pi^2}
\left( {O^{(\beta,\alpha)}_\Lambda(k',\Lambda;p',p)
V^{(\alpha)}_\Lambda(\Lambda,k;p)\over
1-p^2/\Lambda^2}
\right.
+ \left.{V^{(\beta)}_\Lambda(k',\Lambda;p')
O^{(\beta,\alpha)}_\Lambda(\Lambda,k;p',p)\over
1-p'^2/\Lambda^2}
\right) \ ,
\end{eqnarray}
where $V^{(\alpha)}_\Lambda(k',k;p)$ is an effective $NN$-potential 
for a partial wave $\alpha$,
which evolves following the WRG equation for 
a $NN$ potential~\cite{rg1,birse}. 
In $V^{(\alpha)}_\Lambda(k',k;p)$,
$k$ ($k'$) denotes the off-shell 
relative momentum of the two nucleons before
(after) the interaction, $p$ denotes the on-shell momentum, and
$\Lambda$ is the cutoff value specifying the model space.
Some arguments 
in the current operator,
which have been suppressed 
in \Eq{o_me} for simplicity,
are shown. 
We note that, 
because the current operator and $NN$ potential 
are evolved by means of the WRG equation,
the effective current operator acquires 
a dependence on both the initial and final on-shell momenta,
$p$ and $p'$ ($E$ and $E'$),
whereas the effective $NN$-potential acquires 
a dependence on either the on-shell momentum
of the initial state or that of the final state,
$p$ or $p'$ ($E$ or $E'$).
A graphical representation of \Eq{rge} is
shown in Fig.~\ref{fig_rge}.
For an infinitesimal reduction of the cutoff, 
the operator $O$ evolves into 
the renormalized one $O'$
by absorbing the one-loop graph either before 
or after the insertion of the current operator.
The loops include
the intermediate two-nucleon states 
of $\Lambda-\delta\Lambda\le \bar{k}\le\Lambda$,
where $\bar{k}$ is the magnitude of the relative momentum.
The WRG equation given by \Eq{rge} is for a single $NN$ partial wave,
but an extension to a coupled-channel case is straightforward.

One starts with $O=O_{\bar{\Lambda}}$ for a cutoff $\bar{\Lambda}$ and derives
$O_\Lambda$ for $\Lambda (< \bar{\Lambda})$ by solving the WRG equation.
A solution (integral form) 
of the WRG equation, \Eq{rge}, may be written as
\begin{eqnarray}
\eqn{eff_current}
 O_\Lambda^{(\beta,\alpha)} 
&=& \eta \left( O_{\bar{\Lambda}}^{(\beta,\alpha)} 
+ O_{\bar{\Lambda}}^{(\beta,\alpha)} 
\lambda {1\over E-H_{\bar{\Lambda}}^{(\alpha)}} 
\lambda V_{\bar{\Lambda}}^{(\alpha)}
+ V_{\bar{\Lambda}}^{(\beta)} \lambda 
{1\over E'-H_{\bar{\Lambda}}^{(\beta)}} 
\lambda O_{\bar{\Lambda}}^{(\beta,\alpha)}
\right.\nonumber\\
&+& \left. V_{\bar{\Lambda}}^{(\beta)} 
\lambda {1\over E'-H_{\bar{\Lambda}}^{(\beta)}} 
\lambda O_{\bar{\Lambda}}^{(\beta,\alpha)} 
\lambda {1\over E-H_{\bar{\Lambda}}^{(\alpha)}}
\lambda V_{\bar{\Lambda}}^{(\alpha)} \right) \eta \ ,
\end{eqnarray}
where $H_{\bar{\Lambda}}^{(\alpha)}$, $V_{\bar{\Lambda}}^{(\alpha)}$
are the full Hamiltonian and the $NN$-interaction for a partial wave
$\alpha$, respectively, defined in the model space with $\bar{\Lambda}$.
Furthermore,
$\eta$ and $\lambda$ are 
the projection operators 
defined by
\begin{eqnarray}
 \eqn{eta}
\eta &=& \int {\bar{k}^2d\bar{k}\over 2\pi^2} 
\left| \bar{k}\, \right\rangle \left\langle \bar{k}\,
  \right| \,\, ,  \quad \quad \quad
\bar{k}  \leq \Lambda \,\, , \\
\eqn{lambda}
\lambda &=& \int {\bar{k}^2d\bar{k}\over 2\pi^2} 
\left| \bar{k} \,\right\rangle  \left\langle
\bar{k} \, \right| \,\,  ,
\quad \quad \quad
\Lambda < \bar{k} \leq \bar{\Lambda} \,\, ,
\end{eqnarray}
where $\ket{\bar{k}}$ represents the radial part of the free two-nucleon
states with the relative momentum $\bar{k}$.
By inserting \Eq{eff_current} into \Eq{rge},
one can easily check that the integral form of the WRG equation,
\Eq{eff_current}, satisfies 
its differential form, \Eq{rge}.

A nuclear current operator has the properties which a nuclear force
does not possess, {\it i.e.},
the current operator satisfies a (partial) conservation law.
It is interesting to examine the conservation law for the effective
current operator obtained after the model-space reduction.
This subject is discussed in Appendix B.

\vskip 2mm \noindent
{\bf 3. Nuclear current operators}

In this section, we introduce
the nuclear current operators used 
in our numerical renormalization group analysis.
We employ five nuclear current operators
contributing to the GT transition:
two of them are phenomenological models and the others are based on
pionful EFT [\eftpi] with three cutoff
values~\footnote{\label{footnote;Lambda_eft}
The \pieft-based operators contain a Gaussian type cutoff function.
We denote the cutoff by $\Lambda_{eft}$ and distinguish 
it from the sharp cutoff $\Lambda$ introduced in the WRG equation. }.
Each of the five operators
has the same one-body operator,
\begin{eqnarray}
\eqn{one-body}
\vec{A}_1 (\vec{x}) &=& - g_A \sum_{i} 
{\tau^3_i\over 2}\, \vec{\sigma}_i\, \delta^{(3)}(\vec{x}-\vec{r}_i) \ ,
\end{eqnarray}
where $g_A$ is the axial-vector coupling constant, and 
$\vec{\tau}_i$ and $\vec{\sigma}_i$ are 
the isospin and spin operators, respectively, for 
the $i$-th nucleon.
The isospin operator has
only the third component, 
$\tau^3_i/2$, because we consider
the neutral-current reaction,
$\nu d\to \nu pn$, in this work.
It is noted that we neglect the small corrections
from the axial form factor, the induced pseudoscalar current,
and corrections from higher order terms 
({\it e.g.}, $1/M$ corrections) 
because we are concerned with 
the reaction in a low-energy region,
$E_\nu < 20$ MeV, where $E_\nu$ is the 
incoming neutrino energy. 
\begin{figure}
\parbox{.55\textwidth}{\epsfig{file=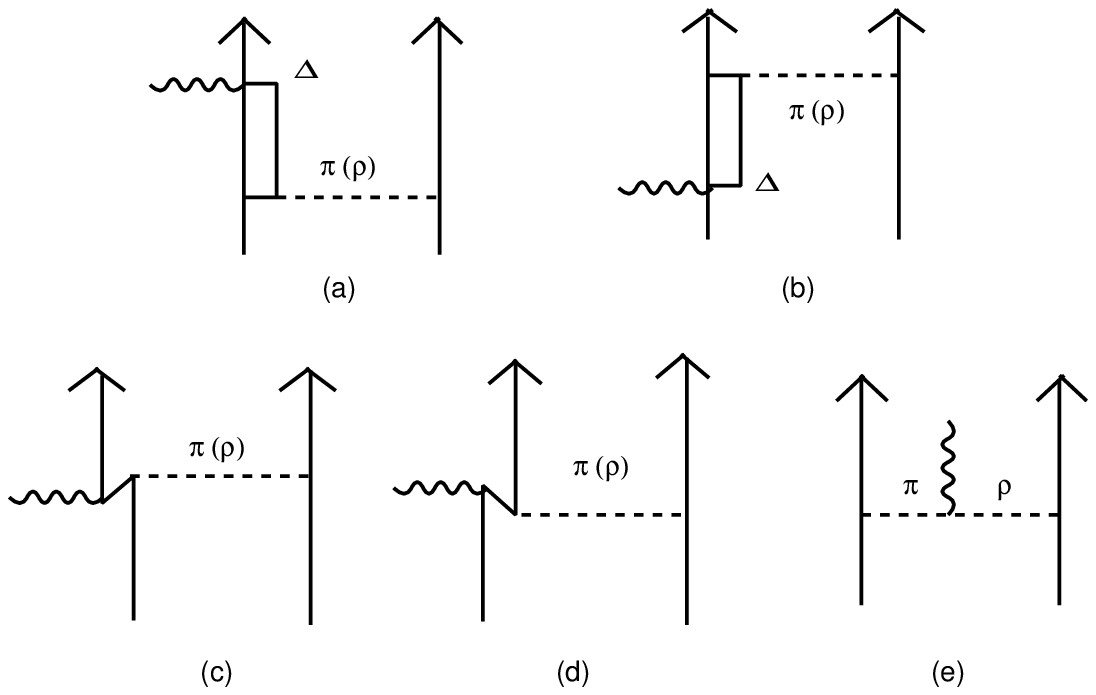,width=0.55\textwidth}
\caption{\label{fig_model} 
Diagrams for axial-vector exchange currents
of phenomenological models. 
Upper two diagrams (a,b) are for the $\Delta$-excitation currents 
induced by the $\pi$ (or $\rho$) exchange.
Lower diagrams (c) and (d) are for the $\pi$ (or $\rho$) 
exchange pair currents,
and diagram (e) is for the $\pi$-$\rho$ current.
}}
\hfill
\parbox{.40\textwidth}{\epsfig{file=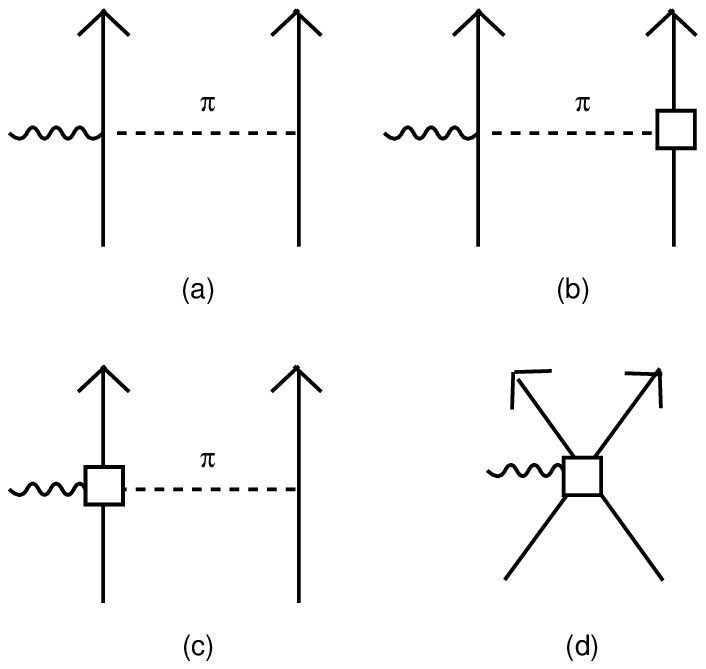,width=0.35\textwidth}
\caption{\label{fig_eft} 
Diagrams for axial-vector exchange currents from \eftpi.
Dashed lines denote pions and 
vertices without (with) white box arise from the leading
(sub-leading) order Lagrangian.
} }
\end{figure}

We now present the two-body operators. 
The operators of two phenomenological models are 
found, {\it e.g.}, in Ref.~\cite{netal}.
The operator for one of the models 
are obtained by calculating five diagrams shown in Fig.~\ref{fig_model}.
The model was originally proposed in Ref.~\cite{schiavilla}
and the strength of the $\Delta$-excitation currents 
in the model was 
fixed to reproduce the experimental tritium $\beta$-decay rate.
We will refer to the model as ``Model I''.
The other model consists of only the $\Delta$-excitation currents
(the upper two diagrams in Fig.~\ref{fig_model}).
The overall strength of the model was adjusted so 
as to reproduce
the same total cross section for $\nu d \rightarrow \nu pn$ as 
that of Model I at the reaction threshold.
We will refer to this model as ``Model II''.
For explicit expressions of the operators, 
see Eqs.~(22)-(26) of Ref.~\cite{netal}.

The nuclear currents based on \eftpi\ are 
obtained 
from heavy-baryon chiral Lagrangian
up to next-to-next-to-next-to leading order 
in Refs.~\cite{ando,ando2}.
Diagrams for the two-body current operator 
are shown in Fig.~\ref{fig_eft} and 
explicit expressions for the 
operators are given in,
{\it e.g.},  Eq.~(20) in Ref.~\cite{ando2}.
We use three \pieft-based operators with different
cutoff values,
$\Lambda_{eft}=500$, 600, and 800 MeV
(see the footnote~\ref{footnote;Lambda_eft}).
For each of the cutoff values,
the coupling constant 
of the axial-vector-four-nucleon contact interaction
shown by diagram (d) in Fig.~\ref{fig_eft} has been fixed 
so as to reproduce the tritium $\beta$-decay 
rate~\cite{petal-prc03}.\footnote{
Recently, it is pointed out in Ref.~\cite{gardestig} that
a contribution
from the contact interaction (Fig.~\ref{fig_eft}(d)) plays
an important role to deduce a value of neutron-neutron
scattering length $a_{nn}$ from experimental data
of the $\pi^-d\to nn\gamma$ reaction.
The authors fixed the coupling
of the contact interaction in a cutoff
independent manner, so as to reproduce a $pp\rightarrow de^+\nu_e$
reaction matrix element from a model calculation.
}

In our numerical analysis, we consider the 
operator, $O$, obtained with
the current operators, $A(\vec{x})$, presented above,
\begin{eqnarray}
\eqn{op_eff}
O = \int d\vec{x}\   e^{i\vec{q}\cdot\vec{x}}
   {\cal A}(\vec{x}) \ ,
\end{eqnarray}
where ${\cal A} (\vec{x})
\equiv A (\vec{x})|_{\vec{R}=0}$,
with $A (\vec{x})$ being the axial-vector current.
The center of mass coordinate of the two nucleons is denoted by $\vec{R}$.
The quantity $\vec{q}$ is the momentum 
transfer to the nuclear current operator.
We will examine the evolution of
the operator $O$ defined in \Eq{op_eff},
rather than the current operator $A(\vec{x})$ itself, using the WRG equation.
Also, a matrix element we calculate is always
that of the operator $O$. 
In what follows, we refer to the operator $O$ in \Eq{op_eff}
also as ``current operator''.

Matrix elements of the operators are calculated
before a model-space reduction ($\Lambda=\infty$),
and the numerical results are shown in Table~\ref{tab_gt_me}.
We calculate contributions (ratio) to 
the matrix elements from the one-body and 
the two-body current operators.
We decompose the contributions from the two-body operators
into those from the deuteron $S$ and $D$-wave states.
For the final neutron-proton scattering state, we consider only the
$^1S_0$ partial wave, which is the dominant state 
in the low-energy region.
The total amplitude from ``Model I'' is normalized to 100\%.
In calculating the matrix elements,
we employ a specific kinematic condition:
$E=-B$, $E'=1$ MeV, and $q=30$ MeV
where $B$ is the deuteron binding energy and
$q=|\vec{q}|$.\footnote{
The total cross section of the
$\nu d \rightarrow \nu pn$ reaction at $E_\nu=20$ MeV
gets a good amount of contribution from the kinematical 
region around this kinematics.}
The wave functions are obtained 
with the Argonne v18 potential\cite{av18}.
Since all of the five operators have
the same one-body operator,
we have 
the same contribution from the one-body operator with the deuteron $S$- and
$D$-waves, 98.87\%.
Meanwhile, the five two-body currents give small contributions
which agree with each other within 1\% accuracy.
We have 1.13, 0.72, 1.42, 1.41, 1.40\% contributions 
from the two-body currents with the deuteron $S$- and $D$-waves
from the Model I and II, and \pieft\ with the three cutoff values
$\Lambda_{eft}=500$, 600, and 800 MeV, respectively. 

\begin{table}[h]
\begin{center}
 \begin{tabular}[t]{clccccc}\hline
$\Lambda$& Model
 & $\braketa{\mbox{1B}}$ &$\braketa{\mbox{2B}}$
 &\multicolumn{2}{c}{$\braketa{\mbox{2B}}$}\\
 (MeV)&&$S+D$ (\%)&$S+D$ (\%)& $S$ (\%)&$D$ (\%)\\ \hline
&Model I          &    98.87&    1.13&     -0.33&     1.46 \\
&Model II         &    98.87&    0.72&      0.00&     0.72 \\
{$\infty$}&
 $\Lambda_{eft}=500$ MeV& 98.87&  1.42&    -1.18&     2.60 \\
&$\Lambda_{eft}=600$ MeV& 98.87&  1.41&    -2.00&     3.41 \\
&$\Lambda_{eft}=800$ MeV& 98.87&  1.40&    -3.26&     4.65 \\ \hline
&Model I          &    97.05&     2.96&     2.85&     0.12 \\
&Model II         &    97.05&     2.56&     2.50&     0.06 \\
{200}&
 $\Lambda_{eft}=500$ MeV& 97.05&  3.25&     3.05&     0.20 \\
&$\Lambda_{eft}=600$ MeV& 97.05&  3.25&     3.04&     0.21 \\
&$\Lambda_{eft}=800$ MeV& 97.05&  3.23&     3.01&     0.22 \\ \hline
&Model I          &    69.22&    30.79&    30.77&     0.03 \\
&Model II         &    69.22&    30.39&    30.36&     0.03 \\
{70}&
 $\Lambda_{eft}=500$ MeV& 69.22& 31.08&    31.05&     0.03 \\
&$\Lambda_{eft}=600$ MeV& 69.22& 31.08&    31.05&     0.03 \\
&$\Lambda_{eft}=800$ MeV& 69.22& 31.06&    31.04&     0.03 \\\hline
 \end{tabular}
\caption{\label{tab_gt_me}
Contributions (ratio) to 
the Gamow-Teller (GT) matrix element, 
$\bra{^1S_0}O\ket{^3S_1(^3D_1)}$, 
from each component of the current operators.
The model space reduction using the WRG equation introduces the sharp
 cutoff and the values of it are shown in the first column.
Each current operator is obtained from
Model I and II, and \pieft\ with $\Lambda_{eft}=500$, 600, 800 MeV,
shown in the second column.
The third and fourth columns 
give the contribution from the one-body (1B) and the two-body (2B) 
operators, respectively.
The 2B contributions are from
the deuteron $^3S_1$ and $^3D_1$ waves, which are separately given in
 the fifth and sixth columns (denoted by $S$ and $D$).
For more details, see the text.
}
\end{center}
\end{table}

We can see from the table that the behaviors of
the two-body currents are quite different.
Model I and II induce
the $^3D_1\to {}^1S_0$ transitions (1.47 and 0.72\%) more strongly
than the $^3S_1\to {}^1S_0$ ones ($-0.33$ and 0.00\%),
because the important operator in the models is the tensor-type
originated from the $\Delta$-excitation currents
exchanging a pion or a rho meson.
In contrast, 
the \pieft-operators generate
the $^3S_1\to {}^1S_0$ transition amplitudes
($-1.18$, $-2.00$, and $-3.26$\% for $\Lambda_{eft}=500$, 600, and 800 MeV,
respectively) and
$^3D_1\to {}^1S_0$ ones
($2.60$, $3.41$, and $4.65$\%),
both of which are larger than those of Model I and II.
The $^3S_1\to {}^1S_0$ transition is 
due to the contact 
interaction, the diagram (d) in Fig.~\ref{fig_eft}, 
whereas the $^3D_1\to {}^1S_0$ transition is due to 
the pion exchange two-body currents,
mainly the diagram (c) in Fig.~\ref{fig_eft}.
We find that a larger cutoff value ($\Lambda_{eft}$) 
leads to a more destructive interference between 
the deuteron $S$ and $D$-wave contributions. 
The interference between the two transitions are so destructive for the
\pieft-based operators and
thus the net contribution is almost the same as those of Model I
and II.

\vskip 2mm \noindent 
{\bf 4. WRG analysis of the current operators}

\begin{figure}[t]
\begin{center}
\includegraphics[width=90mm]{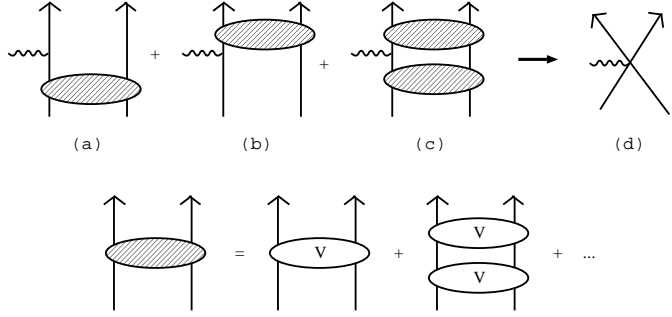}% Here is how to import EPS art
\caption{\label{fig_IA_evolve}
A graphical representation of the renormalization of the effective 
two-body operator due to the one-body operator.
In the diagrams (a)-(c), 
all external two-nucleon systems have the relative momentum
 belonging to the model space while all the intermediate states are the
 high momentum states to be integrated out.
After the model-space reduction, the diagrams (a)-(c) contribute to the
 renormalization of the two-body operator shown by the diagram (d).
}
\end{center}
\end{figure}

Starting with the five sets of 
the nuclear operators discussed in the previous section, 
we calculate effective operators
at the cutoff values $\Lambda = 200$ and 70 MeV 
using the integral form of the WRG equation, \Eq{eff_current}. 
In this work, we define 
the one- and two-body effective currents as
\bea
O^{eff}_1 &=& \eta O_1 \eta\, ,
\\
{O}^{eff}_2 &=& \eta \left[
O_2 + (O_1+O_2) \lambda \frac{1}{E-H}\lambda V
+V\lambda \frac{1}{E'-H}\lambda (O_1+O_2)
\right. \nnb \\ && \left.
+V\lambda \frac{1}{E'-H}\lambda (O_1+O_2)\frac{1}{E-H}\lambda V
\right]\eta \, ,
\label{eq;A2eff}
\eea
where $O_1$ and $O_2$ are the ``bare'' one- and two-body 
current operators, respectively,
which we have already introduced.
We note that 
the high energy parts of the bare one-body current operator $O_1$ 
are included in $O_2^{eff}$,
as graphically shown
in Fig.~\ref{fig_IA_evolve}.  
The cutoffs in the projection operators, $\eta$ and $\lambda$
in \Eqs{eta}{lambda},
are thus $\bar{\Lambda}=\infty$ and $\Lambda=200$
or 70 MeV.

In our WRG analysis,
we observe that the effective two-body operators suddenly jump up
at a momentum around the cutoff.  (We will see it below.)
As shown in Eq.~(\ref{eq;A2eff}) 
(graphically in Fig.~\ref{fig_IA_evolve}),
the bare one-body current operator can give 
a contribution to 
the effective two-body operator,
and this is the origin of the ``jump-up'' structure in the effective 
operators.
In Fig.~\ref{fig_IA_evolve}(a) (Fig.~\ref{fig_IA_evolve}(b)), 
the high momentum states before (after) the insertion of the
current operator are integrated out 
whereas, in Fig.~\ref{fig_IA_evolve}(c),
the high momentum states before and after the insertion of the
current are integrated out to renormalize the two-body operator.
The diagrams (a), (b) and (c) in
Fig.~\ref{fig_IA_evolve} correspond to
the $O_1$ part of the second, third and fourth terms in
Eq.~(\ref{eq;A2eff}), respectively.
Because of the momentum conservation, 
these terms of
Figs.~\ref{fig_IA_evolve}(a)-(c) give non-vanishing contributions
only if the condition, 
$\left|\bar{k}-{q\over 2}\right|\leq \bar{k}^\prime\leq \bar{k}+{q\over
2}$, 
is satisfied; $\bar{k}$ ($\bar{k}^\prime$) is the magnitude of the
relative momentum of the two-nucleon state just before (after) the
insertion of the current operator.
The diagrams, Figs.~\ref{fig_IA_evolve}(a) and (b),
start generating
a large amount of the contributions
at a certain kinematical point.
For example, 
the $O_1$ part of the second term in Eq.~(\ref{eq;A2eff}) 
gives non-vanishing contribution for
\begin{eqnarray}
 \left|\Lambda-{q\over 2}\right|\leq k^\prime\leq \Lambda \ .
\end{eqnarray}
The behavior of the renormalized effective two-body operators 
thus changes suddenly at %a certain 
the value of the momentum
$k$ ($k'$), as discussed above.

\vskip 1mm \noindent
{\bf 4.1 Effective current operators at $\Lambda = 200$ MeV}

Now we are in a position to discuss our result
of the effective operators for $\Lambda=200$ MeV.
We reduce the model space of 
Model I or II or \pieft\ using the
WRG equation,
and evaluate the GT matrix elements
for the kinematics specified before.
The result is shown in Table~\ref{tab_gt_me}.
The contribution from the effective one-body operator, 
97.05\%, is smaller than that (98.87\%) for $\Lambda=\infty$  
by 1.82\%.
This is because a part of the one-body operator 
sensitive to the high momentum part of the wave functions 
is renormalized into the effective two-body 
operator.
In addition, 
we find that most of the two-body $D$-wave contributions
at $\Lambda=\infty$
are also renormalized into 
the two-body $S$-wave ones at $\Lambda=200$ MeV. 
We can see that there still remains some model dependence
among the models and \pieft\ at $\Lambda=200$ MeV. 
On the other hand, the three \pieft-based operators,
which appear differently at $\Lambda=\infty$, give almost the same result at
$\Lambda=200$ MeV.

\begin{figure}[t]
\begin{minipage}[t]{72mm}
\begin{center}
\epsfig{file=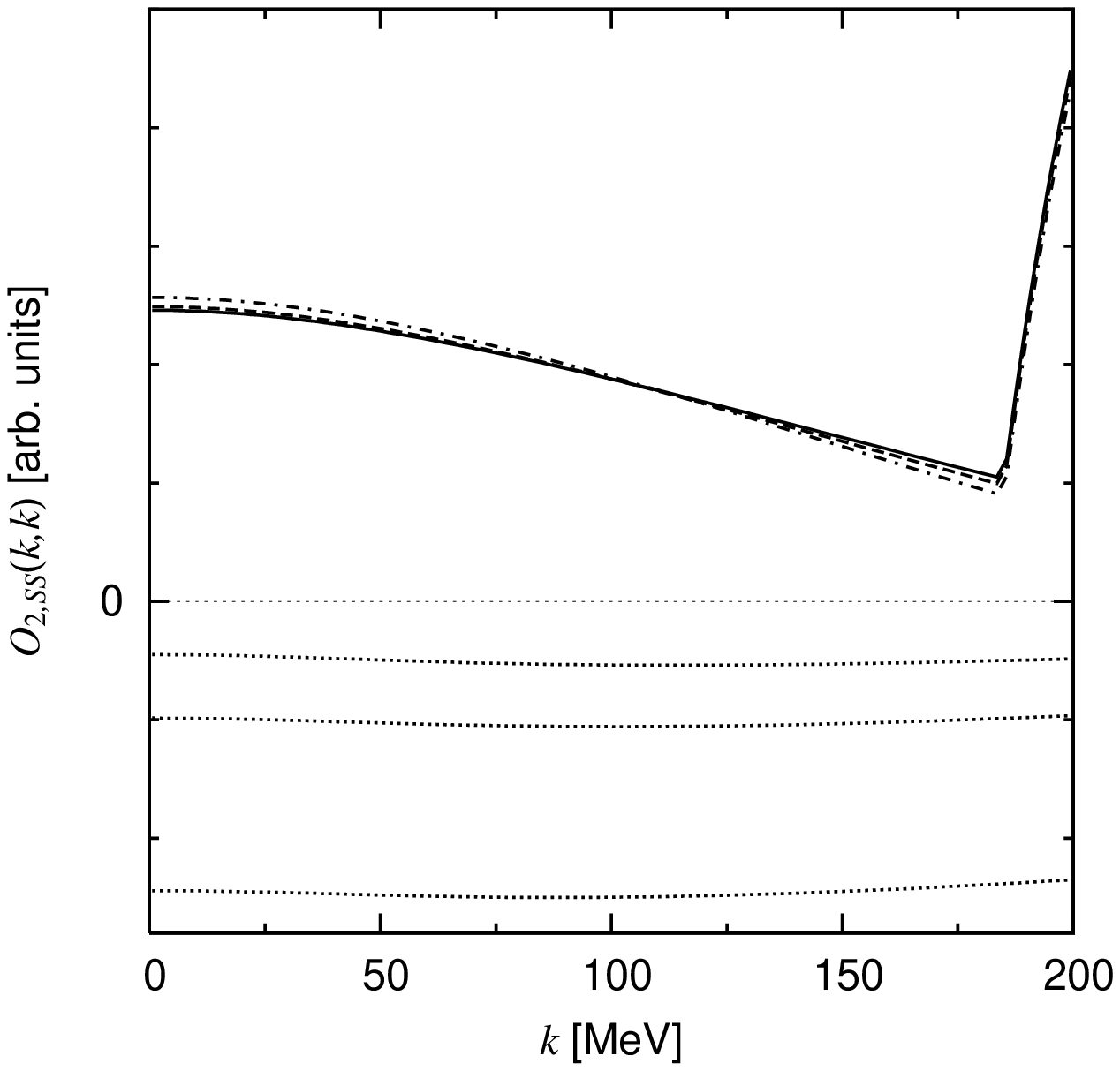,width=0.95\textwidth}
\caption{\label{fig_opeft_ss_200}
Effective two-body operators ($\Lambda = 200$ MeV), 
$O_{2,SS}^{eff}$, for the 
deuteron $S$-wave state from 
\pieft-based $O_{2,SS}$.
The diagonal momentum-space matrix elements are shown.
The lower three dotted curves represent the original $O_{2,SS}$;
from the bottom, $\Lambda_{eft}=800, 600$ and 500 MeV.
The upper three curves represent the $O_{2,SS}^{eff}$.
The solid, dashed and dash-dotted curves are the effective
 two-body currents obtained from \pieft\ with
$\Lambda_{eft}=500, 600$ and 800 MeV, respectively.
}
\end{center}
\end{minipage}
\hspace{10mm}
\begin{minipage}[t]{72mm}
\begin{center}
\epsfig{file=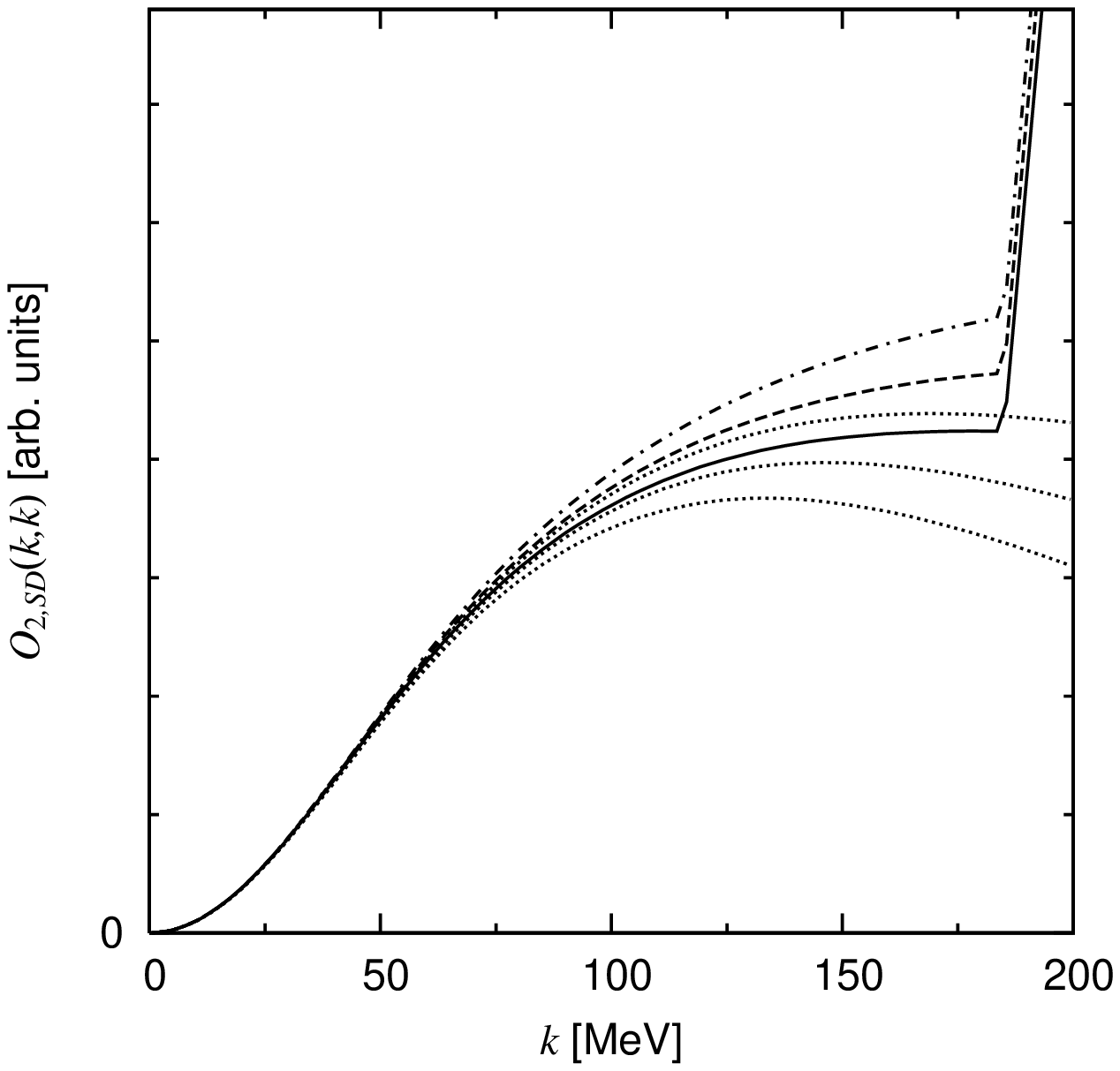,width=0.95\textwidth}
\caption{\label{fig_opeft_sd_200}
Effective two-body operators ($\Lambda = 200$ MeV),
$O_{2,SD}^{eff}$, for 
the deuteron $D$-wave state 
from \pieft-based $O_{2,SD}$.
Three dotted curves represent the original $O_{2,SD}$,
from the bottom, for $\Lambda_{eft}=500, 600$ and 800 MeV.
The other three curves represent the $O_{2,SD}^{eff}$.
The solid, dashed and dash-dotted curves are for 
$\Lambda_{eft}=500, 600$ and 800 MeV, respectively.
Scale of vertical axis is ten times as small as that of
Fig.~\ref{fig_opeft_ss_200}.
The other features are the same as those in Fig.~\ref{fig_opeft_ss_200}.
}
\end{center}
\end{minipage}
\end{figure}

In Figs.~\ref{fig_opeft_ss_200} and \ref{fig_opeft_sd_200},
we show the radial part of effective two-body current
operators for $\Lambda=200$ MeV 
obtained from the \pieft-based operators. 
We denote the effective operator for
the initial deuteron $S$- and $D$-wave states
(and the final $^1S_0$ state) by
$O_{2,SS}^{eff}(k',k)$ and $O_{2,SD}^{eff}(k',k)$, respectively.
We plot the diagonal part of them,
{\it i.e.}, $k'=k$, and also that of the ``bare'' current 
operators, $O_{2,SS}(k,k)$ and $O_{2,SD}(k,k)$.
We find that the curves of the effective operators in 
the figures suddenly change at 
$k=\Lambda-q/2=185$ MeV (``jump-up'' structure).
This is because, as discussed above, at this point 
the diagrams Figs~\ref{fig_IA_evolve}(a) and (b)
start contributing to the effective two-body operators.
We also find that the cutoff $\Lambda_{eft}$ dependence 
among the original \pieft-based operators at $\Lambda=\infty$
are much less observed at this resolution of the system.
However, the disappearance of the model dependence
is not as perfect as the case for the nuclear force,
$V_{low\mbox{-}k}$, in which the model dependence of phenomenological nuclear
forces are almost not observed at
$\Lambda(=2.1 \mbox{\rm fm}^{-1}) \simeq 400$ MeV\cite{rg1,vlowk}.
We find that there still remains a significant model dependence
($\Lambda_{eft}$-dependence)
for $\Lambda \simeq$ 400 MeV (we did not show it), and that the model
dependence of the central-type \pieft-based
nuclear current almost disappears at $\Lambda=200$ MeV
(Fig.~\ref{fig_opeft_ss_200}).
Furthermore, we observe
the model dependence in the tensor-type operator
even at $\Lambda = 200$ MeV
(Fig.~\ref{fig_opeft_sd_200}).
This result has been expected because all of the operators
have the same pion-range mechanism.
Since the \pieft-based operators consist of the 
same one-pion-exchange currant and the contact currents,  
they look almost the same only
when we observe the operators with a resolution insensitive to
details of mechanisms other than the one-pion-exchange mechanism.
A resolution of $\Lambda\sim 400$ MeV is still sensitive to 
the short-range mechanisms,
while that of $\Lambda\sim 200$ MeV is not. 

\begin{figure}[t]
\begin{minipage}[t]{72mm}
\begin{center}
 \epsfig{file=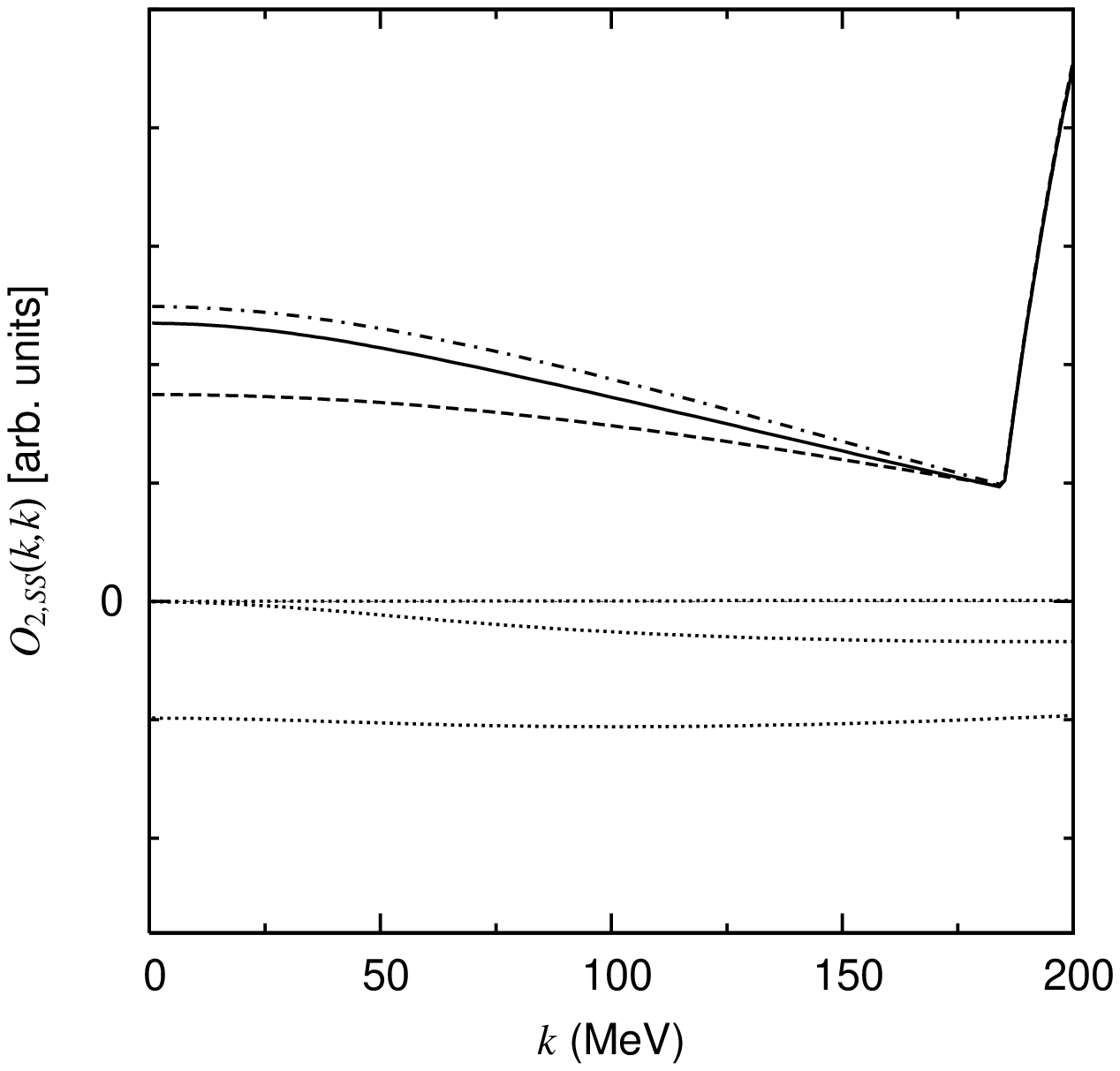,width=0.95\textwidth}
 \caption{\label{fig_op_ss_200}
 Effective two-body operator $O_{2,SS}^{eff}$ ($\Lambda=200$ MeV)
 for the deuteron $S$-wave state from various models of
 $O_{2,SS}$.
 The diagonal momentum-space matrix elements are shown.
 The lower three dotted curves represent the bare $O_{2,SS}$;
 from the bottom, \pieft, Model I and II.
 We show \pieft\ for $\Lambda_{eft}=600$ MeV.
 The dotted curve for Model II is almost on the zero axis.
 The upper three curves represent the $O_{2,SS}^{eff}$ 
 and the solid, dashed and dash-dotted curves are the effective
 two-body currents obtained from Model I, II and \pieft.
 }
\end{center}
\end{minipage}
\hspace{10mm}
\begin{minipage}[t]{72mm}
\begin{center}
\epsfig{file=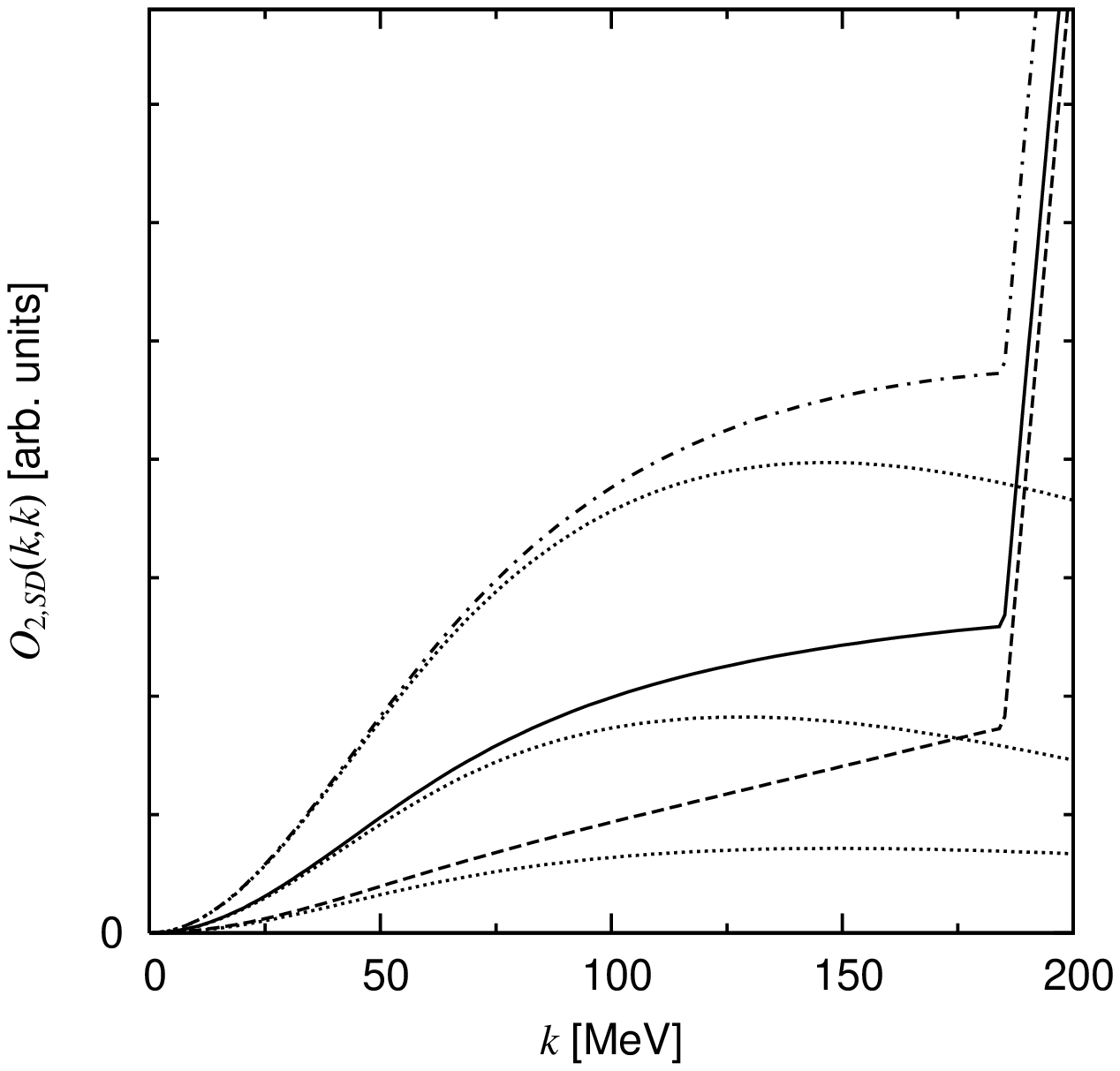,width=0.95\textwidth}
\caption{\label{fig_op_sd_200}
Effective two-body operator $O_{2,SD}^{eff}$ 
($\Lambda = 200$ MeV)
for the deuteron $D$-wave
from various models of $O_{2,SD}$.
The three dotted curves represent the bare $O_{2,SD}$;
from the bottom, Model II, Model I and \pieft.
The other three curves represent the $O_{2,SD}^{eff}$.
The solid, dashed and dash-dotted curves are the effective
 two-body currents obtained from Model I, Model II and \pieft.
Scale of the vertical axis is ten times as small as that of
Fig.~\ref{fig_op_ss_200}.
The other features are the same as those in Fig.~\ref{fig_op_ss_200}.
}
\end{center}
\end{minipage}
\end{figure}
In Figs.~\ref{fig_op_ss_200} and \ref{fig_op_sd_200},
we plot the diagonal part of the effective two-body 
operators 
at $\Lambda=200$ MeV
from the phenomenological models 
for the initial $S$- and $D$-wave deuteron states,
respectively.
We also plot the ``bare'' operators, 
$O_{2,SS}(k,k)$ and $O_{2,SD}(k,k)$ from the models and \pieft\
in the figures.
We find in these figures
that the model dependence still clearly 
remains at this scale.
This is because the starting operators are model dependent 
even on the one-pion exchange mechanism.

\vskip 1mm \noindent 
{\bf 4.2 Effective current operators for $\Lambda = 70$ MeV}

We now discuss our result of the WRG analysis 
at $\Lambda=70$ MeV. 
In Table~\ref{tab_gt_me},
we present our result for the ratio of the amplitudes
from the one- and two-body operators.
We find a significantly reduced contribution (69.22\%) 
from the effective one-body operator at $\Lambda=70$ MeV
because the cutoff value is now rather small
and thus the large portion of the contribution from the bare one-body
operator is renormalized into the effective two-body operators. 
We also find that most of the contribution from the
tensor-type two-body operator is renormalized
into the central-type two-body operator at $\Lambda=70$
MeV\footnote{In a rather small model space,
the tensor nuclear force also plays an unimportant role,
which is discussed in Appendix C.
}.
\begin{figure}[t]
\begin{minipage}[t]{72mm}
\begin{center}
\epsfig{file=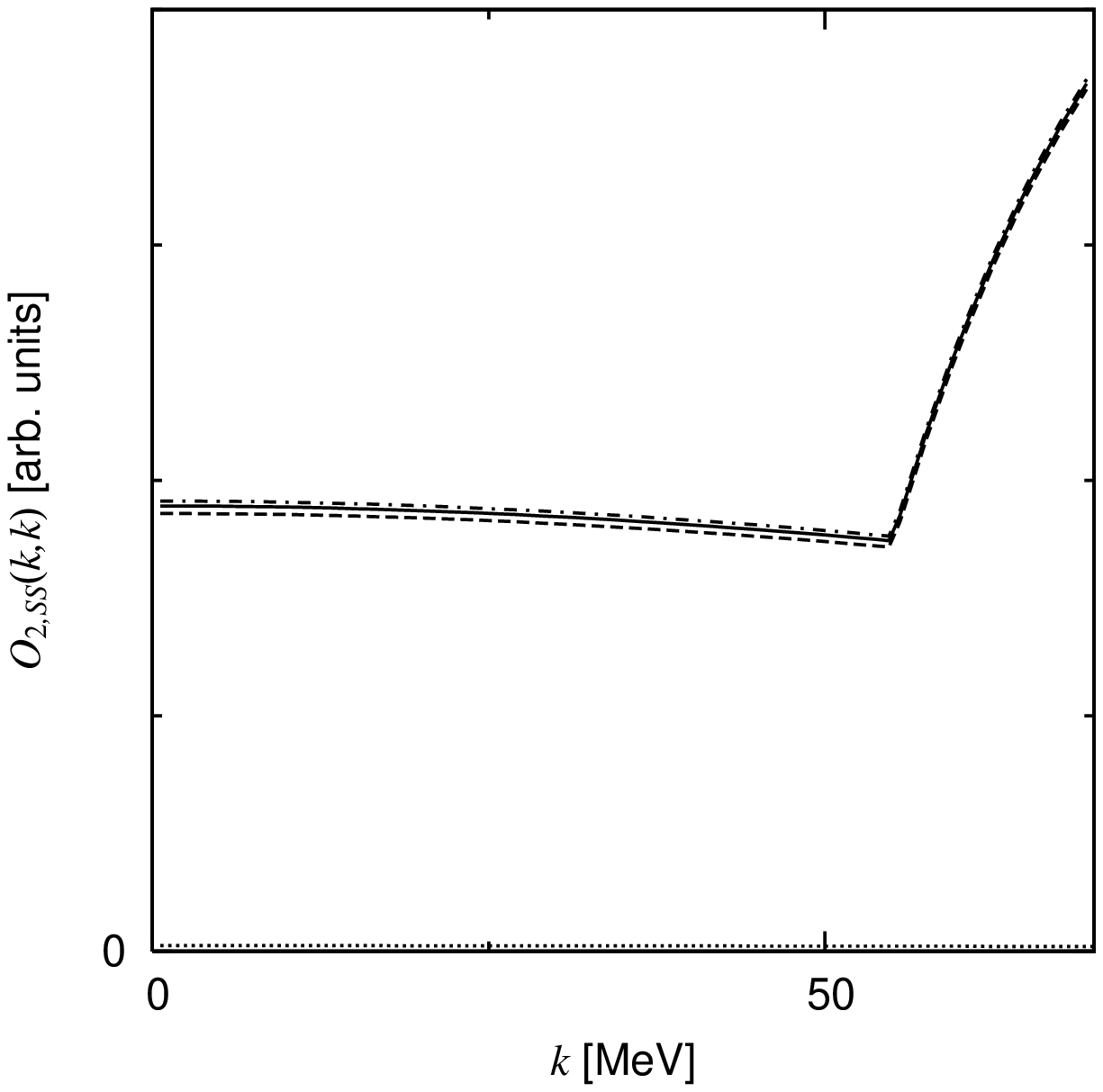,width=0.95\textwidth}
\caption{\label{fig_op_ss_70}
Effective two-body axial-vector current operator $O_{2,SS}^{eff}$
($\Lambda = 70$ MeV) for the initial deuteron $S$-wave.
The dotted curve (almost on the zero line) is 
the same as the solid curve in Fig.~\ref{fig_op_ss_200}.
The other three curves represent the effective currents for
$\Lambda$ = 70 MeV.
The solid, dashed and dash-dotted curves are respectively
derived from Model I, Model II and \pieft\
with $\Lambda_{eft}=600$ MeV.
}
\end{center}
\end{minipage}
\hspace{10mm}
\begin{minipage}[t]{72mm}
\begin{center}
\epsfig{file=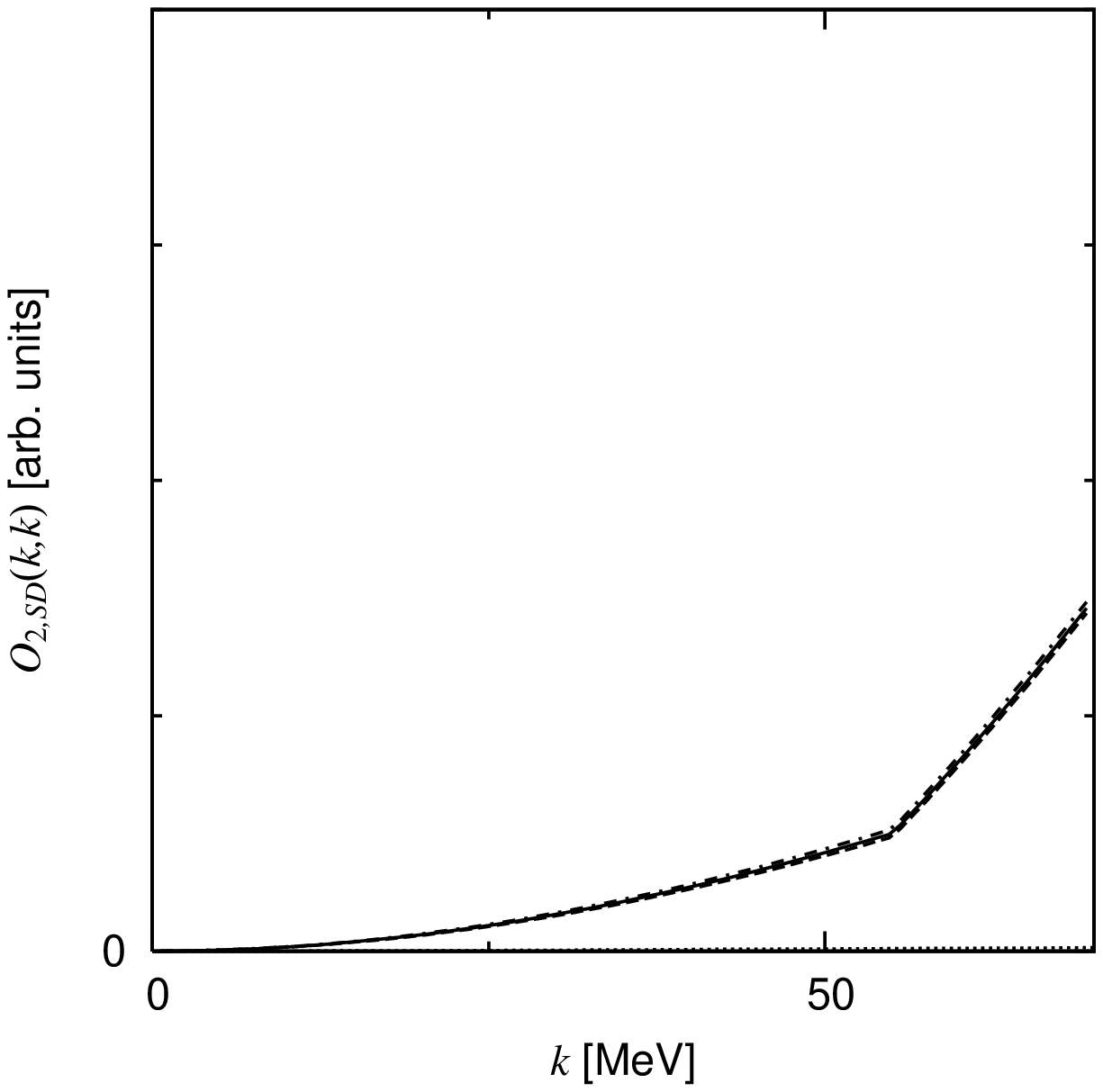,width=0.95\textwidth}
\caption{\label{fig_op_sd_70}
Effective two-body axial-vector current operator $O_{2,SD}^{eff}$
($\Lambda= 70$ MeV) for the initial deuteron $D$-wave.
The dotted curve (almost on the zero line) is 
the same as the solid curve in Fig.~\ref{fig_op_sd_200}.
Scale of the vertical axis is 
ten times as small as that of
Fig.~\ref{fig_op_ss_70}.
The other features are the same as 
those in Fig.~\ref{fig_op_ss_70}.
}
\end{center}
\end{minipage}
\end{figure}
In Figs.~\ref{fig_op_ss_70} and \ref{fig_op_sd_70},
we show our results of the effective operators 
at $\Lambda=70$ MeV for the initial deuteron 
$S$- and $D$-wave states, respectively.
We find again that the curves of the effective 
two-body operators suddenly change 
at $k=\Lambda-q/2=55$ MeV, as discussed earlier,
due to the renormalization from the bare one-body operator.
Since the three \pieft-based operators become to be very similar to
each other already at $\Lambda=200$ MeV, 
we show an evolution of one ($\Lambda_{eft}=600$ MeV) of them. 
With this resolution of the system,
the dependence on modeling
the details of the small scale physics 
is not seen any more.
Therefore, the three different nuclear operators give essentially 
the same GT matrix elements in the kinematical region 
within the model space of $\Lambda=70$ MeV.

We mention here the on-shell energy dependence of the effective nuclear
current operator.
As discussed in the section 2, 
an effective nuclear current obtained with the WRG equation [\Eq{rge}] 
acquires a dependence 
on both the initial and final on-shell energies.
In our case, the on-shell energy for the initial state (the deuteron) is
fixed. Therefore, we examined the dependence of the effective two-body
current on the on-shell energy of the
final scattering state.
We found a significant dependence on the on-shell energy for
this model space ($\Lambda=70$ MeV), which indicates an importance of
the on-shell energy dependence of a NEFT-based operator defined in a small
model space.

\vskip 1mm \noindent 
 {\bf 4.3 \piless-based current operator from
 effective current operator}

\begin{figure}[t]
\begin{center}
\epsfig{file=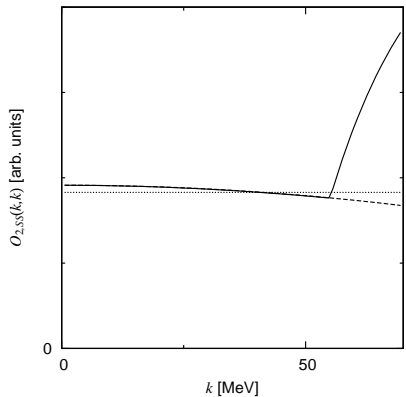,width=0.475\textwidth}
\caption{\label{fig_op_ss_70_fit}
The simulation of the effective two-body axial-vector 
current with $\Lambda=70$ MeV using the
 contact current operators.
The solid curve is the same as the solid curve in
 Fig.~\ref{fig_op_ss_70}.
The dotted curve simulates the solid curve using the contact current
 operator with no derivative. The dashed curve additionally has the
 contact current with two derivatives.
}
\end{center}
\end{figure}

We simulate the obtained low-momentum current operator with
$\Lambda=70$ MeV using the \piless-based 
parameterization~\cite{bck,bc-plb01}
\begin{eqnarray}
O^{\nopix}_{2,SS}(k',k) &=& L_{1A} + {K_{1A}\over 2} (k^2 + k'^2) \ .
\end{eqnarray}
It is clear that the part 
of the effective operator
after $k=\Lambda-q/2$
cannot be well simulated by this parameterization.
We therefore simulate an effective operator
which does not include
the contributions from the diagrams, Figs.~\ref{fig_IA_evolve}(a) and (b).
The result of the simulation is shown in Fig.~\ref{fig_op_ss_70_fit}.
Even the one-parameter fit is fairly good, and
the two-parameter fit yields an almost perfect simulation.
We note that this contact current, whose parameters have been
fixed by fitting to the diagonal components, also simulates well the
off-diagonal components of the effective current operator.
The \piless-based current operator can be obtained from the models or
\pieft\ in this way, and this may be understood as a relation between them.

Even \piless-based current operator 
was determined from the models (or \pieft) 
using the WRG equation, 
it is still missing 
the contributions from the diagrams of
Figs.~\ref{fig_IA_evolve}(a) and (b).
Because these contributions are more (less)
for a larger (smaller) momentum transfer,
validity of the operator is limited to a small momentum transfer region.
For the $\nu d \rightarrow \nu pn$ reaction at $E_\nu=20$ MeV
and $T_{NN}(=E')=1$ MeV, the allowed region of the momentum transfer is
$15\ltap q \ltap 30$ MeV. 
The contributions from the diagrams of
Figs.~\ref{fig_IA_evolve}(a) and (b)
to the GT matrix
elements are 1.8\%, 2.8\% and 5.1\% for $q=15, 20$ and 30 MeV,
respectively.
When a few percent level precision 
is required, this contribution is significant.
Although there is still a kinematical region where the omission of
the ``jump-up'' part is a good approximation, 
the \piless-based current operator constructed 
in the previous paragraph does not accurately predict 
the $\nu d\to\nu pn$ reaction cross sections
in the solar neutrino energy regime ($E_\nu\ltap 20$ MeV).
One useful idea to overcome the difficulty 
can be found in the calculations of 
More Effective Effective Field Theory (MEEFT) (or EFT*)~\cite{kp-arnps04},
which is discussed in the next paragraph.

MEEFT has been successfully applied to calculations 
of reaction rates for electroweak processes in few-nucleon 
system (see Ref.~\cite{kp-arnps04} and references therein), 
and is essentially equivalent to the proper NEFT as discussed 
in Ref.~\cite{meeft}.
In the MEEFT calculation of a matrix element, 
the bare one-body current operator is sandwiched by wave 
functions obtained with a phenomenological nuclear force.
The high-momentum components of the bare one-body operator
have neither been integrated out nor been renormalized into 
the effective two-body operator.\footnote{
A phenomenological nuclear force has a cutoff
typically larger
than a cutoff introduced in a NEFT-based nuclear force.
Strictly speaking, therefore, (very) small effects of high momentum
components of
the one-body operator outside the model space should be
renormalized into
the effective two-body operator, even in the case in which we use a
phenomenological nuclear force. 
}
Meanwhile, the cutoff ($\Lambda_{eft}$) is introduced only in the two-body
operators.
If we follow the manner of MEEFT,
we would not have the ``jump-up'' structure in the effective two-body currents.
Thus we can safely parameterize the effective two-body operator 
using the NEFT-based operator and use it in a wider kinematical region.

\vskip 2mm \noindent
{\bf 5. Conclusion}

In this work,
we extended our previous renormalization group analysis 
of the nuclear force to the nuclear current operator.
We derived the Wilsonian renormalization group (WRG) equation
for the current operator and used it for 
the analysis of the operators.

In our RG analysis,
we studied the five 
nuclear axial-vector currents, 
which are from the two phenomenological models and 
\pieft\ with the three cutoff values,
associated with 
the GT matrix element for the $\nu d\to\nu pn$ reaction
in the solar neutrino energy regime.
The original operators are different from one another due to the
difference in describing the details of the small scale physics;
even the pion range mechanism is model dependent.
In spite of the difference, 
these operators give essentially the same GT
matrix element in the low-energy region.
It was found that,
as the model space is reduced,
the model dependence
gradually disappears and 
an essentially unique effective operator is
obtained at the small model space with $\Lambda=70$ MeV.
It was noted that the three \pieft-based operators converge to a single
operator in the relatively larger model space with $\Lambda=200$ MeV
because they have the same pion range mechanism.
As a result,
the differences among the original operators 
in the reaction mechanism,
such as 
the deuteron $D$-wave contribution in the matrix element of the
two-body operator, disappear.
Our RG analysis clearly showed
the reason why the nuclear current operators of 
the phenomenological models and
those of NEFT give essentially the
same electroweak reaction rate. The reason is that
as long as reactions with the kinematics included in the small
model space are concerned, the reactions 
hardly probe the details of the
small scale physics which make a difference among these nuclear
operators.

We simulated the single effective operator obtained with the WRG equation
using the \piless-based parameterization.
A good simulation indicates that if we
observe the operators of \pieft\ or phenomenological models roughly enough,
they look like that of \piless, not only in the strong sector but also
in the electroweak sector\footnote{
If the pion range mechanism of the phenomenological nuclear current
models were the same as that of \pieft, we could have also connected
them through WRG, as we did for the nuclear force in Ref.~\cite{rg1}.
}; this is the relation among these operators.
Furthermore, we can regard 
the \piless-based current operator 
as model-independent because
it 
simulates well the unique, model-independent effective
current operator defined in a certain small model space.
However, the accurate simulation can be done only if we omit
the ``jump-up'' part of the effective operators.
This omission limits ourselves to a rather small kinematical
region (small momentum transfer) where the omission gives negligible
effects to the GT matrix element.
In this regard,
we discussed the advantage of using MEEFT to avoid the problem.

Finally, we make some remarks here.
Although we performed the RG analysis of the nuclear current operator
for the $\nu d\rightarrow \nu pn$ reaction, the result we obtained
should be the case for other electroweak processes in
two-nucleon system such as the
$np\rightarrow d\gamma$ reaction~\cite{pmr-prl95,cs-prc99} and 
the $pp\rightarrow de^+\nu_e$ 
reaction~\cite{schiavilla,petal-prc03,bc-plb01,pkmr-aj98,kr-npa99}.
That is, in nuclear current operators for those
two-nucleon processes,
one can also find the relation between
models and NEFT using the WRG method.
The operators for the $\nu d\rightarrow \nu pn$ reaction are studied as
an example to study the general relation.
With the general result, we mention
an impact of this work on non-nuclear physics
communities, such as neutrino physics and astrophysics.
Theoretical cross sections
of various reactions in two-nucleon system
have been often used by people in these communities
as input parameters in their
analysis of their own problems.
Although the agreement in the theoretical cross sections between the
different approaches (models, NEFT) has been a good basis for those
people to rely on the theoretical values, the situation is not totally
satisfactory because
the reason for the agreement has not been clearly shown.
Now we can provide the communities with a firmer basis,
{\it i.e.},
the equivalence
relation between phenomenological models and NEFT from the viewpoint of
RG, which we have demonstrated in this work.

\vskip 2mm \noindent
{\bf Acknowledgments}

This work is supported by the Natural Science 
and Engineering Research Council of Canada.
SA is also supported by  Korean Research Foundation
and The Korean Federation of Science and Technology 
Societies Grant funded by Korean Government
(MOEHRD, Basic Research Promotion Fund). 

\vskip 2mm \noindent

{\bf Appendix A: Derivation of Wilsonian renormalization equation for nuclear
 current operator}

We start with a matrix element, evaluated in the momentum space, of an
operator $O$ defined in a model space:
\begin{eqnarray}
\eqn{op_me_app}
\braketa{O} = \int_0^\Lambda\! {k^2\! dk\over 2\pi^2} \int_0^\Lambda\! {k'^2\! dk'\over 2\pi^2}\;
\psi^{(\beta)*}_{p'}(k')\; O^{(\beta,\alpha)}_\Lambda(k',k;p',p)\; \psi^{(\alpha)}_p(k) \ ,
\end{eqnarray}
where $\psi^{(\alpha)}_p(k)$ ($\psi^{(\beta)}_{p'}(k')$)
is the radial part of
the wave function for the initial (final) two-nucleon state, and
$\alpha$ ($\beta$) specifies the partial wave.
The quantities $k$ and $k'$ ($p$ and $p'$) are
the magnitudes of the (on-shell) relative momenta of the two-nucleon
system.
The cutoff $\Lambda$, which is the maximum magnitude of the relative momenta,
specifies the size of the model space for the two-nucleon states.
The radial part of the current operator between the $\beta$ and $\alpha$
partial waves is denoted by
$O^{(\beta,\alpha)}_\Lambda(k',k;p',p)$, which may depend on the
on-shell momenta and the cutoff.

We differentiate the both sides of \Eq{op_me_app} with respect to
$\Lambda$, and obtain
\begin{eqnarray}
\eqn{eq_diff1}
{\partial\braketa{O}\over \partial\Lambda} &=& 
\int_0^\Lambda\! {k^2\! dk\over 2\pi^2} {\Lambda^2\over 2\pi^2}\;
\psi^{(\beta)*}_{p'}(\Lambda)\; O^{(\beta,\alpha)}_\Lambda(\Lambda,k;p',p)\; \psi^{(\alpha)}_p(k)\nonumber\\
&+& {\Lambda^2\over 2\pi^2}\int_0^\Lambda\! {k'^2\! dk'\over 2\pi^2} \;
\psi^{(\beta)*}_{p'}(k')\; O^{(\beta,\alpha)}_\Lambda(k',\Lambda;p',p)\; \psi^{(\alpha)}_p(\Lambda)\nonumber\\
&+& \int_0^\Lambda\! {k^2\! dk\over 2\pi^2} \int_0^\Lambda\! {k'^2\! dk'\over 2\pi^2}\;
\psi^{(\beta)*}_{p'}(k')\; {\partial\,
O^{(\beta,\alpha)}_\Lambda(k',k;p',p)\over\partial\Lambda}\; \psi^{(\alpha)}_p(k) \ .
\end{eqnarray}
Here we write the wave function as
\begin{eqnarray}
\eqn{eq_wave}
\psi^{(\alpha)}_p(k) = 2\pi^2{\delta(p-k)\over k^2} 
+ {1\over E_p - E_k + i\epsilon}
\int_0^\Lambda\! {k'^2\! dk'\over 2\pi^2} V^{(\alpha)}_\Lambda(k,k';p)\;
\psi^{(\alpha)}_p(k') \ ,
\end{eqnarray}
where $E_x\equiv x^2/M$ with $M$ being the nucleon mass.
The quantity $V^{(\alpha)}_\Lambda(k,k';p)$ is the $NN$-potential
defined in the model space specified by $\Lambda$; the
$\Lambda$-dependence of $V^{(\alpha)}_\Lambda(k,k';p)$ is
controlled by the WRG equation for 
$NN$ potential~\cite{rg1,birse}. 
It is noted, in the case of the deuteron wave function,
that the first term in the r.h.s. of \Eq{eq_wave} does not exist,
and that the normalization is such that the amplitudes are the same as
those of the wave function from $V_{\Lambda =\infty}$ whose squared
integral is normalized to be unity.
Using \Eq{eq_wave}, we can rewrite \Eq{eq_diff1} as
\begin{eqnarray}
\eqn{eq_diff2}
{\partial\braketa{O}\over \partial\Lambda} &=& 
\int_0^\Lambda\!\! {k^2\! dk\over 2\pi^2}
\int_0^\Lambda\!\! {k'^2\! dk'\over 2\pi^2}\,
{\Lambda^2\over 2\pi^2}\;
\psi^{(\beta)*}_{p'}(k')\; V^{(\beta)}_\Lambda(k',\Lambda;p')\;
{1\over E_{p'} - E_\Lambda}\;
O^{(\beta,\alpha)}_\Lambda(\Lambda,k;p',p)\; \psi^{(\alpha)}_p(k)\nonumber\\
&+& \int_0^\Lambda\!\! {k^2\! dk\over 2\pi^2}
\int_0^\Lambda\!\! {k'^2\! dk'\over 2\pi^2} \,
{\Lambda^2\over 2\pi^2}\;
\psi^{(\beta)*}_{p'}(k')\; O^{(\beta,\alpha)}_\Lambda(k',\Lambda;p',p)\; 
{1\over E_{p} - E_\Lambda}\;
V^{(\alpha)}_\Lambda(\Lambda,k;p)\;
\psi^{(\alpha)}_p(k)\nonumber\\
&+& \int_0^\Lambda\!\! {k^2\! dk\over 2\pi^2} \int_0^\Lambda\!\! {k'^2\! dk'\over 2\pi^2}\;
\psi^{(\beta)*}_{p'}(k')\; {\partial\,
O^{(\beta,\alpha)}_\Lambda(k',k;p',p)\over\partial\Lambda}\; \psi^{(\alpha)}_p(k) \ .
\end{eqnarray}
Now we impose a condition
that the matrix element is invariant with respect to changes in
$\Lambda$, {\it i.e.}, 
\begin{eqnarray}
\eqn{eq_condition}
{\partial\braketa{O}\over \partial\Lambda} = 0\ ,
\end{eqnarray}
then we find from \Eq{eq_diff2} the Wilsonian renormalization group
equation for the operator $O$:
\begin{eqnarray}
\eqn{eq_rg2}
{\partial\, O^{(\beta,\alpha)}_\Lambda(k',k;p',p)\over\partial\Lambda} &=&
{M\over 2\pi^2} V^{(\beta)}_\Lambda(k',\Lambda;p')\,
{\Lambda^2\over \Lambda^2-p'^2}\,
O^{(\beta,\alpha)}_\Lambda(\Lambda,k;p',p)\nonumber\\ 
&+& {M\over 2\pi^2}\, O^{(\beta,\alpha)}_\Lambda(k',\Lambda;p',p)\,
{\Lambda^2\over \Lambda^2-p^2}\,
V^{(\alpha)}_\Lambda(\Lambda,k;p) \ .
\end{eqnarray}
Note that \Eq{eq_rg2} is a sufficient condition of \Eq{eq_condition}, but not a
necessary condition. Nevertheless, \Eq{eq_rg2} is physically appealing because
this equation can also be derived with the path integral method, as
stated in footnote \ref{footnote;path_int}, without concerning about whether \Eq{eq_rg2} is a
necessary condition or not.

\vskip 2mm \noindent
{\bf Appendix B: Conservation law for effective current operator}

It is interesting to study a conservation law for an effective
current operator obtained with the WRG equation [\Eq{rge}].
Suppose that we start with a model in which the (vector or axial-vector)
current operator $J^\mu$ satisfies an equation
\begin{eqnarray}
\eqn{current_eq}
 q_\mu J^\mu &=& S \ ,
\end{eqnarray}
where $S$ denotes a source term and $q^\mu$ is the momentum transfer
from the external current to a nuclear system.
(We will use $q^\mu$ as the momentum transfer in the following.)
We also suppose that this model is defined in a model space whose size
is specified by a cutoff $\bar{\Lambda}$.
Now we reduce the model space of the current operator using the WRG
equation up to a cutoff $\Lambda$. We obtain
\begin{eqnarray}
\eqn{eff_current-appen}
 J^\mu_{eff} 
=\eta \left( J^\mu + J^\mu \lambda {1\over E-H} \lambda V 
+ V \lambda {1\over E'-H} \lambda J^\mu
\right.
+ \left. V \lambda {1\over E'-H} \lambda J^\mu \lambda {1\over E-H}
     \lambda V \right) \eta \ ,
\end{eqnarray}
where $J^\mu_{eff}$ is the effective current operator defined in the
model space with $\Lambda$.
The full Hamiltonian and the $NN$-interaction are respectively denoted
by $H$ and $V$, both of which are defined in the model space with
$\bar{\Lambda}$.
The quantities $E$ and $E'$ are the kinetic energies of the two-nucleon
system in the initial and final states, respectively.
The projection operators $\eta$ and $\lambda$ are defined by
\begin{eqnarray}
\label{eq;eta}
\eta &=& \int {d^3 \bar{k}\over (2\pi)^3} \left| \bar{\vecbox{k}}\, \right\rangle \left\langle \bar{\vecbox{k}}\,
  \right| \,\, ,  \quad \quad \quad
\left| \bar{\vecbox{k}} \, \right| \leq \Lambda \,\, , \\
\label{eq;lambda}
\lambda &=& \int {d^3 \bar{k}\over (2\pi)^3} \left| \bar{\vecbox{k}} \,\right\rangle  \left\langle
\bar{\vecbox{k}} \, \right| \,\,  ,
\quad \quad \quad
\Lambda < \left| \bar{\vecbox{k}} \, \right| \leq \bar{\Lambda} \,\, ,
\end{eqnarray}
where $\ket{\bar{\vecbox{k}}}$ represents the free two-nucleon states with the
relative momentum $\bar{\vecbox{k}}$.

Now, with the \Eqs{current_eq}{eff_current-appen},
the equation which the effective current satisfies is
\begin{eqnarray}
\eqn{current_eff_eq}
 q_\mu J^\mu_{eff} &=& S_{eff} \ ,
\end{eqnarray}
with
\begin{eqnarray}
\eqn{eff_source}
 S_{eff} 
= \eta \left( S + S \lambda {1\over E-H} \lambda V 
+ V \lambda {1\over E'-H} \lambda S
\right.
+ \left. V \lambda {1\over E'-H} \lambda S \lambda {1\over E-H}
     \lambda V \right) \eta \ .
\end{eqnarray}
Therefore, if the current $J^\mu$ in the starting model is
conserved ($S=0$ in \Eq{current_eq}), then the effective current 
$J^\mu_{eff}$ is also conserved ($S_{eff}=0$ in \Eq{current_eff_eq}).
As has been done in the text, we simulated the effective current with the
use of the contact currents.
In such a case, the contact currents violate \Eq{current_eff_eq} to a
certain extent.
However, as we have seen, the simulation is quite accurate, up to the
``jump-up'' part, and therefore the violation of the equation is very small.

\vskip 2mm \noindent
{\bf Appendix C: Tensor nuclear force in pionless effective field theory}

In describing low-energy two-nucleon system, 
two types of nuclear effective field theories (NEFTs) have been often
used.
One of them considers the nucleon and the pion explicitly [\eftpi] while
the other treats only the nucleon as an explicit degree of freedom [\piless].
Although both NEFTs successfully describe low-energy $NN$ system,
there is a distinct difference between them in dealing with the tensor force.
In \pieft\ with a cutoff regularization scheme, the tensor force which
comes from the one-pion-exchange potential (OPEP) is considered to be
a leading order (LO) term, and is iterated to all orders by solving
the Schr\"odinger equation.
Thus, the tensor force in \pieft\ is LO and treated
non-perturbatively\footnote{
Although a perturbative treatment of the OPEP was
proposed in the KSW counting scheme\cite{pds}, 
it turned out that the OPEP has to be treated
non-perturbatively\cite{pds2}.
}.
In contrast, treatments of the tensor force in \piless\ are as follows.
In the cutoff scheme, the tensor force is a next-to-leading order (NLO) term
and is treated non-perturbatively (iteration to all orders).
In the KSW counting scheme\cite{pds}, the tensor force is a
next-to-next-to-leading order (N$^2$LO) term and is perturbatively
treated.
Thus, the tensor force is considered to be less important
in \piless, which is based on the counting rules used.
However, it is still interesting to examine whether the tensor force
indeed has different importance in \pieft\ and \piless.
This examination can be done quantitatively
by deriving \vpiless\ from \vpi, for which
the renormalization group plays an essential role
(\vpi\ (\vpiless) is the nuclear force based on \eftpi\ (\piless)).
This is what we discuss in this appendix, using a result given in
our previous paper\cite{rg1}.

We use \vpi\ and \vpiless\ presented in Ref.~\cite{rg1} in the analysis
here. More specifically, we use \vpiII\ which 
consists of the OPEP plus contact interactions with zero, two and
four derivatives , and \vpilessII\ which 
consists of contact interactions with zero, two and
four derivatives.
The cutoff values, which specify the size of the model space for the
nuclear forces, are $\Lambda=200$ MeV for \vpiII\ and
$\Lambda=70$ MeV for \vpilessII.
In Ref.~\cite{rg1},
the parameterization of these nuclear forces is given in Eq.~(2.14) and
Appendix B, and the numerical values of the parameters are given in
Table I.
The two interactions, \vpiII\ and \vpilessII\ , are related to each other
through the Wilsonian renormalization group equation (Eq.~(2.3) of the
reference).

Now we focus our attention on the $^3S_1$-$^3D_1$ partial wave, because
we are interested in the importance of the tensor force.
We calculate the $NN$ scattering phase shifts and the deuteron binding
energy using \vpiII\ and \vpilessII.
In order to study the importance of the tensor force,
we switch on and off the tensor force included in \vpiII\ and
\vpilessII. The result is shown in Table~\ref{tab_piless}.
We can observe that the tensor force in \vpilessII\ gives only a tiny
effect to the observables.
On the other hand, the tensor force in \vpiII\ plays an unnegligible
role, as expected.\footnote{
However, the effect of the tensor force in \vpiII\ is not so
significant. This is because we used the model space which is rather
small for \pieft. We adopted the small model space so that the
two-pion-exchange potential can be well simulated by the contact
interactions; we consider, for simplicity, only the OPEP as a mechanism
including the pion explicitly.
Adopting a larger model space makes the role of the tensor force more
significant.
As is well known, the omission of the tensor force in a phenomenological
nuclear force leads to no bound state in the $^3S_1$-$^3D_1$ partial wave.
}
In this way, we conclude that the tensor force has different importance
in \pieft\ and \piless, which supports the counting rules used so far.

\begin{table}[h]
\caption{\label{tab_piless}
The importance of the tensor force in $NN$ observables.
The first column indicates the nuclear force used, while the second one
 specifies whether the tensor force contained in the nuclear force is
 turned on or off.
The third (fourth) column presents the phase shifts of the $NN$
$^3S_1$-$^3D_1$ partial wave scattering for $p=10 (30)$ MeV;
$p$ is the on-shell relative momentum of the two nucleons.
The fifth column gives the deuteron binding energy.
}
\begin{center}
  \begin{tabular}[t]{ccccc}\hline
   &tensor force &$\delta^{(^3S_1)}_{p=10 {\rm MeV}}$
  &$\delta^{(^3S_1)}_{p=30 {\rm MeV}}$&B.E. \\\hline
 \vpiII&ON& 164.47& 137.19&  2.225\\
 ($\Lambda=200$ MeV)&OFF& 163.46& 135.15&  1.835\\\hline
 \vpilessII&ON& 164.47& 137.35&  2.223\\
 ($\Lambda=70$ MeV)&OFF& 164.46& 137.34&  2.217\\\hline
 \end{tabular}
\end{center}
\end{table}

\end{document}